# Precisely aligned graphene grown on hexagonal boron nitride by catalyst free chemical vapor deposition

Shujie Tang,[1,] Haomin Wang,[1, a)] Yu Zhang,[2] Ang Li,[1] Hong Xie,[1] Xiaoyu Liu,[1] Lianqing Liu,[2] Tianxin Li,[3] Fuqiang Huang,[4] Xiaoming Xie,[1,a)] Mianheng Jiang[1]

[1] State Key Laboratory of Functional Materials for Informatics, Shanghai Institute of Microsystem and Information Technology, Chinese Academy of Sciences, 865 Changning Road, Shanghai 200050, P.R. China

2 State Key Laboratory of Robotics, Shenyang Institute of Automation, Chinese Academy of Sciences,114 Nanta Street, Shenhe District, Shenyang 110016, P.R. China

[3]National Laboratory for Infrared Physics, Shanghai Institute of Technical Physics, Chinese Academy of Sciences, 500 Yu Tian Road, Shanghai 200083, P.R.China

[4]CAS Key Laboratory of Materials for Energy Conversion, Shanghai Institute of Ceramics, Chinese Academy of Sciences, Shanghai, 200050, P.R. China

a) Correspondence and requests for materials should be addressed to: hmwang@mail.sim.ac.cn, xmxie@mail.sim.ac.cn

TEL:+86-21-62511070 ext 3403    FAX:  +86-21-62524732

**Abstract:** To grow precisely aligned graphene on h-BN without metal catalyst is extremely important, which allows for intriguing physical properties and devices of graphene/h-BN hetero-structure to be studied in a controllable manner. In this report, such hetero-structures were fabricated and investigated by atomic resolution scanning probe microscopy. Moiré patterns are observed and the sensitivity of moiré interferometry proves that the graphene grains can align precisely with the underlying h-BN lattice within an error of less than $0.05°$. The occurrence of moiré pattern clearly indicates that the graphene locks into h-BN via van der Waals epitaxy with its interfacial stress greatly released. It is worthy to note that the edges of the graphene grains are primarily oriented along the armchair direction. The field effect mobility in such graphene flakes exceeds 20,000 $cm^2 \cdot V^{-1} \cdot s^{-1}$ at ambient condition. This work opens the door of atomic engineering of graphene on h-BN, and sheds light on fundamental research as well as electronic applications based on graphene/h-BN hetero-structure.

**Introduction:**

Graphene has attracted enormous interest and intense research activities due to its rich physical properties in fundamental research and versatile potentiality to revolutionize many applications, especially in electronics and opto-electronics.[1, 2] As the charge carriers in graphene are very sensitive to the underlying substrates, such as the surface roughness,[3] charge impurity,[4-6] surface phonon energy,[7] and the adsorbates,[8] long coherent length and high mobility of charge carriers are usually achieved on mechanically exfoliated graphene that is normally either suspended [9, 10] or transferred on h-BN.[11, 12] As h-BN has a flat surface free of dangling bonds and charge impurities, the carrier scattering in graphene/h-BN can be greatly reduced. Compared to suspended graphene, graphene on h-BN has a physical support to avoid the mechanical collapsing, thus facilitates the subsequent processing. For these reasons, single crystal h-BN is believed to be an ideal substrate for graphene-based devices. Recent experiment showed that the weak periodic potential imposed by a supper-lattice on graphene/h-BN can generate extra Dirac cones in its energy dispersion. [13] Anderson localization was realized in ultra-high-quality graphene sandwiched by h-BN via tuning the carrier density[14] and in the similar way a high on/off ratio of $10^6$ was obtained in graphene transistor through h-BN barriers.[15] Moreover, theoretical calculation predicts that AB stacked graphene/h-BN structure could generate a tunable band gap.[16] However, graphene/h-BN devices in above experiments were fabricated by the direct transfer technology, which usually gives rise to a random orientation of graphene, not to mention its extremely low yield. The random stacking always leads to the uncertainty in the electronic properties of graphene/h-BN. Controllable synthesis of well-aligned graphene/h-BN is thus highly demanding.

Several reports on fabrication of graphene/h-BN hetero-structure are already available based on metal catalytic CVD process[17, 18], the process is quite similar to that of catalytic graphene growth on metal substrates[19-22], however, the unavoidable transfer process added to the complexities to such approaches and lowered their significances. Thus far, only a few experiments were reported on the CVD growth of graphene on h-BN without metal catalysts.[23-25] Our earlier work has demonstrated 200 nm single crystal graphene grains on h-BN flakes.[24] It was found that graphene nucleates mainly at the defective sites of substrate. Carbon atoms accumulate around the nucleus and grow continuously into graphene domains. However, the detailed structures of the graphene domains and h-BN are rarely investigated. Especially, the graphene lattice registration to h-BN is yet unknown. To precisely measure the atomic feature of graphene, one may have to turn to scanning tunneling microscopy (STM), which is normally performed under ultrahigh vacuum environment and requires the sample to be conductive. As the h-BN substrate is insulating, scanning atomic force probe microscopy (AFM) provides an alternative, regardless of the conductivity of substrate and the height variation modulated from the local density of states.[26-28] Here we use atomic force microscope (AFM) to investigate the CVD-grown hetero-epitaxy of graphene on h-BN. The appearance of moiré interference pattern confirms that the graphene grains can align themselves precisely with the underlying h-BN substrate under optimized processes. This work opens the door for atomic engineering of graphene on h-BN, shedding light on many intriguing possibilities for fundamental research as well as device application on the graphene/h-BN hetero-structure.

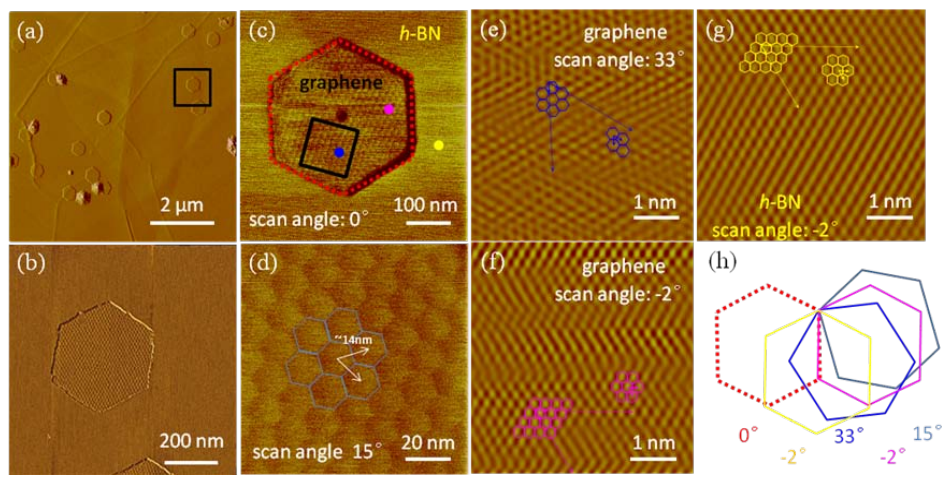

**Figure 1** Determination of the rotational orientation of monolayer graphene with respect to h-BN. (a) A typical topography of the graphene grains on h-BN surface. Almost all the grains are in hexagonal shape with the opposite sides parallel. The hexagons are aligned on the whole h-BN surface. It clearly indicates anisotropic growth of graphene; (b) A zoom-in view from the black box in panel (a). A sizable superstructure with a periodicity much larger than the lattice constant of both graphene and h-BN was observed on graphene; (c) Friction image of a single crystal graphene on h-BN. Dashed red line marks the edges of this grain; (d) A closer view ($100 \times 100$ nm$^2$) in the black box of (c), the superstructure exhibits hexagonal symmetry with lattice constant about 14 nm; Regular hexagons are superimposed on the images to demonstrate the giant lattice. The friction images (e), (f) and (g) show the atomic lattice of $5 \times 5$ nm$^2$ area taken from the blue, pink and orange dot areas in panel(c), regular hexagons demonstrate the lattice of both graphene and h-BN, respectively. Zigzag directions (in dashed line) and lattice vectors (in solid line) are indicated by arrows. The atomic images are filtered to improve clarity, and the corresponding raw

images of **e**–**g** are shown in supplementary materials. The small distortion in images **e** and **f** was due to slight slip in the movement of the AFM tip. The scanning angle of the tip is sometimes adjusted in order to obtain clear images with sample location fixed; (h) All the model hexagons with different scan angles in previous panels are rescaled for comparison. The hexagons are all parallel after correcting the scan angle.

**Results:**

The graphene/h-BN samples were grown by dissociating methane ($CH_4$) at a pressure of 12 mbar around 1200 °C, as described in our earlier report.[24] Fig. 1a shows a typical AFM image of h-BN surface after CVD growth. Many single-layer graphene grains (see Fig. 1a) are formed in hexagonal shape on h-BN, typically with their opposite edges parallel. Their edges exhibit similar orientation signifying an epitaxial growth. A zoom-in view (Fig. 1b) reveals a sizable superstructure with the periodicity much larger than the lattice constants of both graphene and h-BN. A detailed investigation is followed to understand the geometric relation between graphene grain, lattice and the enormous periodicity. In order to reduce the influence of tip effect, graphene grains of moderate size are chosen as depicted in Fig.1c and 1d. The superstructure has a hexagonal symmetry with lattice constant about 14 nm.

In order to obtain the direct evidence for the correlation of the crystal structure and orientation, atomic-resolution friction measurements on the different locations of graphene/h-BN sample are carried out, and the friction images are shown in Figure 1e, 1f and 1g, respectively. Unless otherwise stated, all atomic-resolution friction images in the following part are taken in an area of $5 \times 5$ nm$^2$. Fig. 1e and 1f, show the atomic lattice of the hexagonal graphene grain while Fig. 1g shows that of h-BN. Zigzag directions (in dashed line) and lattice vectors (in solid line) are indicated by arrows. The angle between the two zigzag directions is ∼ 60 °. Regular hexagons are superimposed on the images to demonstrate the lattice of graphene and h-BN. The characteristics of honeycomb lattice in both graphene and h-BN can be clearly observed. Figure 1h presents the representative models of hexagon in Figure 1c (grain hexagon with the scan angle 0°), 1d (superstructure hexagon with the scan angle 15°), 1e (graphene lattice hexagon with scan angle 33°), 1f (graphene lattice hexagon, with the scan angle -2°), and 1g (h-BN lattice hexagon with the scan angle -2°). The model hexagons with different scanning angles are rescaled for comparison. The fact that they are all parallel (after correcting the scan angle) verifies the epitaxy of graphene on h-BN from atomic level.

Normally there are two most energetically favorable edge configurations in graphene: zigzag and armchair.[29-37] As the neighboring zigzag and armchair edges sharing the same corner are exclusively 30°/90°/150° apart, the hexagonal graphene grains with 120° corners are solely armchair- or zigzag-edged. By comparing the hexagonal models in Fig. 1e and 1f with that in Fig.1c, it is easy to find that the edges of graphene grain on *h*-BN are all along the armchair direction. Of course, the edge configuration is not microscopically clear because of the difficulty to obtain atomic resolution on the edge via friction AFM. Usually, zigzag edges are preferred in hexagonal graphene flakes grown on metal.[19-21, 38] Although the mechanism is still under debate, the catalytic metal surface is believed to play an important role.[38-42] In addition, zigzag edges are always observed in graphene after metal-assisted anisotropic etching [43-45] and hydrogen plasma etching [46]. Hydrogen partial pressure is very important in determining the configuration of graphene edges due to its multiple roles in edge reconstruction, and etching of graphene domains, as well as removal of surface-adsorbed C atoms.[47] A well-shaped, regular hexagonal hole with zigzag edges can also be obtained on $SiO_2$-supported graphene via oxidation at 500 °C.[33, 48] These experiments indicate the superior thermal stability of zigzag edge to armchair one in etching process. Different from etching process, the formation of graphene edges involves higher partial pressure of carbon source. Density function calculations predict that armchair edges have lower formation energy than zigzag edges in the absence of metal surface,[49] supported by the observation of armchair edges in graphene islands formed on SiC (0001) [34]. It is believed that the higher density of states near the zigzag edge leads to higher formation energy while the armchair edge is free of boundary states.[49] Our findings seem to be consistent with this scenario.

The hexagonal superstructure in Figure 1b-d is believed to be moiré pattern whose periodicity depends on the lattice mismatch between graphene and *h*-BN, as reported in earlier scanning tunneling microscopic studies on mechanically exfoliated graphene.[13,50,51] Typical moiré pattern in graphene exhibits hexagonal symmetry and a periodicity of about 14 nm (as shown in Fig. 1d), and aligns with the underlying *h*-BN. The existence of moiré pattern gives us two strong indications. Firstly, in the conventional epitaxy, the top layer lattice is bonded and firmly locked to the substrate lattice, the film-substrate interplay results in an interfacial stress depending on the lattice mismatch and growth condition. Here in our case, the interfacial stress between graphene and the *h*-BN is relaxed, and the growth continues according to a so called van der Waals epitaxy. The substrate affects the hetero-epitaxial process only through the van der Waals (vdW) interaction.[52, 53] it is necessary to mention that the film-substrate interaction is adequate to line up two lattices, but too weak to introduce crystallographically constrained epitaxy. Secondly, from the moiré periodicity one can determine precisely the lattice orientation between graphene and *h*-BN as we will do in the following section.

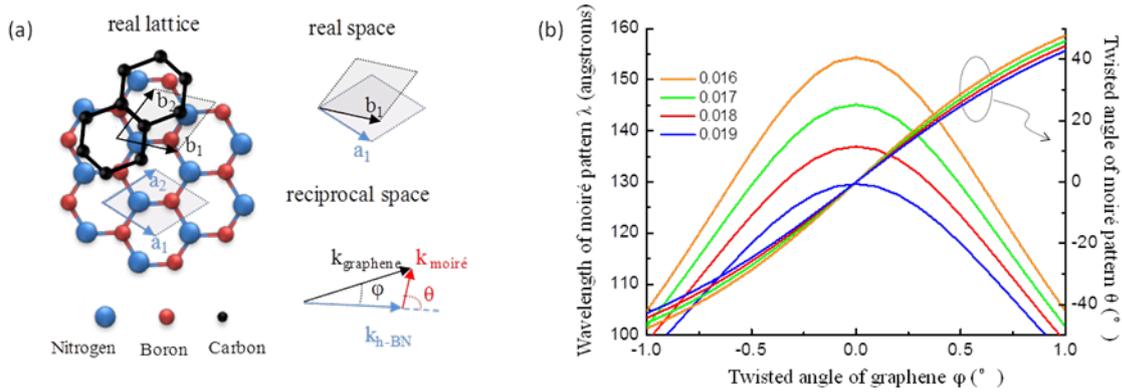

**Figure 2** Characteristics of moiré pattern on graphene/*h*-BN. (a) Representation of a rotated monolayer graphene on *h*-BN surface. The honeycomb lattice, real lattice vectors and the corresponding reciprocal lattice vectors of both graphene and *h*-BN are plotted respectively. In the reciprocal space, moiré pattern vector $k_{Moiré}$ connects the *h*-BN reciprocal lattice vector $k_{h-BN}$ to the graphene reciprocal lattice vector $k_{graphene}$, $\varphi$ is the rotation of $k_{graphene}$ with respect to $k_{h-BN}$, $\theta$ describes the relative rotation of $k_{moiré}$ with respect to $k_{h-BN}$; (b) Moiré pattern wavelength and its rotation angle with respect to the *h*-BN as a function of mis-orientation angle between graphene and *h*-BN (several lattice mismatch values are presented)

**Discussion:**

The schematic in Fig. 2a is an explanation to moiré geometry. When *h*-BN is overlaid with a graphene layer, new symmetry may emerge leading to the moiré pattern. Its orientation and wavelength are fixed and can be derived mathematically. As shown in Fig. 2a, the hexagonal lattice of *h*-BN is defined by vectors $(a_1, a_2)$, where $a_1$ and $a_2$ can be written as $a_1 = a(\frac{\sqrt{3}}{2}, -\frac{1}{2})$ and $a_2 = a(\frac{\sqrt{3}}{2}, \frac{1}{2})$, respectively. $a$ represents the lattice constant of *h*-BN. Similarly, graphene lattice is represented by $(b_1, b_2)$. The lattice of graphene is mis-oriented counterclockwise by an angle $\varphi$ from the *h*-BN and shorter in length by a factor of $\delta$. The factor $\delta$ denotes lattice mismatch between $h$-BN and graphene. Then the reciprocal lattice vector can be described as $k_{h-BN} = \frac{2\pi}{a}(1,0)$ for $h$-BN and $k_{graphene} = \frac{2\pi}{a(1-\delta)}(\cos\varphi, \sin\varphi)$ for graphene, respectively. The reciprocal lattice vectors of the moiré pattern is given by: $k_{moire} = k_{graphene} - k_{h-BN}$. Therefore, the moiré pattern in real space has a wavelength $\lambda = \frac{2\pi}{|k_{moire'}|} = \frac{(1-\delta)a}{\sqrt{2(1-\delta)(1-\cos\varphi) + \delta^2}}$ and an mis-oriented angle $\theta = \tan^{-1}\frac{\sin\varphi}{\cos\varphi + \delta - 1}$ to $h$-BN. Here we use 2.51 angstroms as the value of $a$. The wavelength of moiré pattern (λ) and its orientation relative to the $h$-BN (θ) as functions of mis-orientation angle between graphene and $h$-BN (φ) are plotted in Fig. 2b with lattice mismatch (δ) in the range 1.6% to 1.9%. (A more generalized analysis is presented in Fig. S6 and the subsequent description). It is found that the wavelength of moiré pattern decreases with the increase of the absolute value of δ. It reaches maximum when φ equals to zero. In the other words, moiré pattern in the maximal dimension can be observed when graphene grain aligns perfectly with the underlying $h$-BN. The estimated δ of 1.6~1.9% yields λ ranging from 13 to 15 nm, which is in good agreement with the measured value. We noticed that λ decays very fast with the increase of φ. When φ exceeds 5º, the moiré lattice constant drops to a value of less than 3 nm, beyond the limit of AFM (See supplementary materials Fig. S6d). Furthermore, the alignment of moiré pattern (θ) strongly depends on how much graphene is mis-orientated from $h$-BN (φ). A nonzero φ as small as 0.5º can cause a rotation of more than tens of degree for the moiré pattern. As the measurement error for lattice orientation is less than ±3 ° in our experiments (See Supplementary Materials content 4), the mis-orientation between graphene and $h$-BN should be less than 0.05 °. That is to say, the graphene lattice is precisely aligned with that of the $h$-BN substrate.

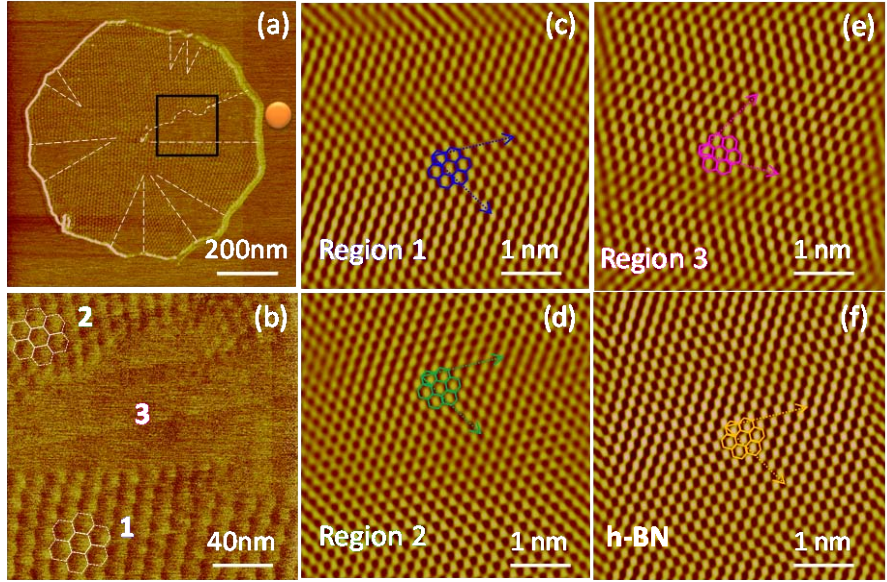

**Figure 3** Friction imaging of polycrystalline monolayer graphene on $h$-BN. (a) A typical polycrystalline graphene grown on $h$-BN. The graphene flake with giant moiré patterns is stitched with several domains where the moiré pattern is invisible. Dashed lines depict the shapes of the "blank" regions. The blank regions seem radiated from the center of the graphene flake; (b) Magnified friction images from the black box in panel (a). The moiré patterns exhibit discontinuity. Regular hexagons demonstrate the lattice of moiré pattern; (c), (d), (e) and (f) are the atomic scans (5 × 5nm$^2$) taken from the blue, pink, green and orange positions in panel (a) and (b), regular hexagons demonstrate the lattice of graphene and *h*-BN, respectively. Zigzag directions (in dashed line) and lattice vectors (in solid line) are indicated by arrows. The atomic images are filtered to improve clarity. Above images are measured with the same scan angle. In region "1" and "2", their moiré patterns have a 12° angular difference although the graphene lattice in both regions seems to follow that in *h*-BN substrate pretty well. The graphene lattice in region "3" is rotated of about 30° with respect to the *h*-BN lattice. The corresponding unfiltered raw images of **c**–**f** are shown in Fig. S7.

Additional information about the domain orientation comes from polycrystalline monolayer graphene on *h*-BN. The polycrystalline graphene can be encountered in samples grown at slightly lower temperature. As shown in Fig. 3a, the graphene flake is stitched with several domains where moiré patterns are missing (at least invisible to AFM). The moiré pattern shows discontinuity across different regions. It's obvious in region "1" and "2" while blank in region 3 as shown in a zoom-in view (Fig. 3b). A careful examination on region "1" and "2" found that their moiré patterns have a 12° angular difference although the graphene lattice in both regions seems to follow that in substrate pretty well (Fig. 3c,d and f). Recall from Fig. 2b, it is obvious that the moiré pattern rotation angle (θ) varies very sensitively on the lattice mis-orientation angle (φ). A rotation of moiré pattern by 12° corresponds to 0.2° graphene mis-orientation from h-BN, too small to be noticed by our AFM. On the contrary, the graphene lattice in region "3" is at 30° to the *h*-BN that causes a drop of moiré wavelength down to about 0.5 nm (See Fig. S6c and S6d), which is almost undetectable. We also noticed that the moiré patterns in neighboring domains are usually oriented away from each other by a large amount, typically near 30º (as shown Fig. 3a). This characteristic angle may correspond to a second lowest energy for graphene registry to *h*-BN, similar to the graphene/graphene stacking.[54]. This phenomenon differs clearly from the earlier results of polycrystalline graphene grown on transitional metals where random lattice orientations were found among domains.[55-57] In all, lower temperature causes the growth of polycrystalline graphene in which the lattices are imperfectly aligned with the *h*-BN substrate. The large-angle grain boundaries may indicate a metastable state for graphene/*h*-BN epitaxy.

It is noticed that defective sites are often found near the center of graphene flakes. This finding confirms our early conclusion that these sites are nucleation centers.[24] The carbon adatoms have extremely high mobility on *h*-BN at elevated temperatures hence only the defective sites or step edges can trap them to form nuclei. Compared with the inert surface of graphene, sites with dangling bonds are much more reactive to incoming atoms or fragments, consequently the growth of graphene continues laterally after nucleation.

The transport properties of graphene devices were measured with low-frequency lock-in technique. The field effect mobility extracted from two graphene samples is above $2\times10^4 cm^2 \cdot V^{-1} \cdot s^{-1}$ at ambient condition. The high values indicate that the carrier scattering is not significant at the graphene/*h*-BN interface. The preserved high mobility in graphene on the other hand proves that the interaction between graphene and *h*-BN is weak, further supporting our observation of epitaxy via van der Waals force.

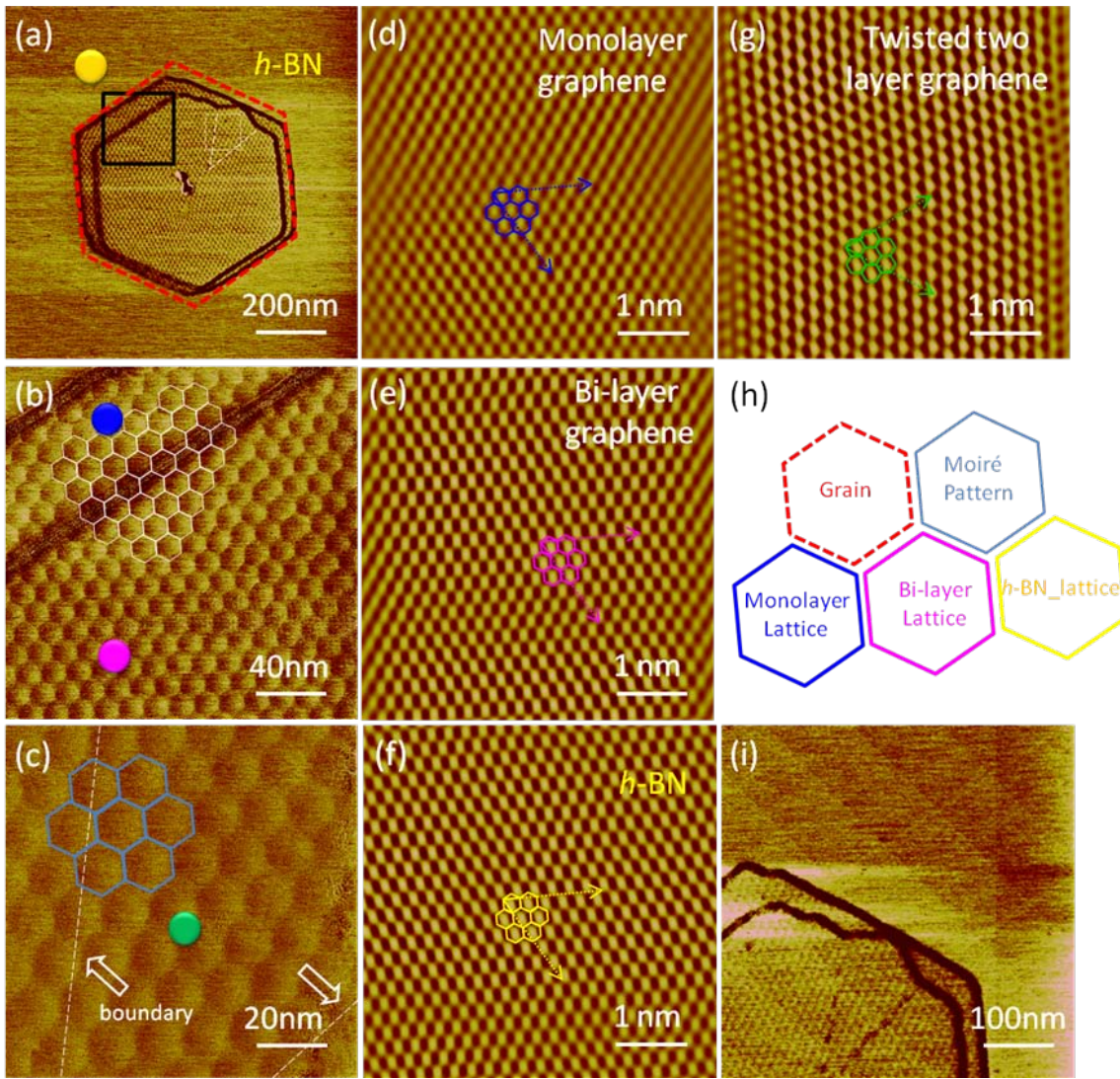

**Figure 4** Investigation of the morphology of bilayer graphene grown on h-BN. (a) A typical bilayer graphene on h-BN. Both the top and bottom layer of the bilayer are in hexagonal shape with the opposite sides parallel. The nearly perfect alignment between top- and bottom-layer hexagons can be observed. Dashed red line marks the grain edges of the bottom layer. To be noted that the right-up part of the top layer marked between two white dashed lines is a graphene domain with different lattice orientation; (b) Magnified friction image from the black box in panel (a) is shown. Moiré pattern on both terraces exhibits uninterrupted periodicity across the step edge; (c) Magnified friction image from the white box in panel (a) is shown. The dashed white lines represent the tilted interfaces of different domains in the top layer. Continuous moiré pattern was also observed on the graphene domains stitched together at tilted interfaces. Regular hexagons demonstrate the patterns; The friction images (d), (e), (f) and (g) show the (scan 5 nm ×5 nm) atomic lattices taken from the blue, pink, orange and green dots in panel(a), (b) and (c), some regular hexagons demonstrate the lattice of graphene and h-BN, respectively. Zigzag directions (in dashed line) and lattice vectors (in solid line) are indicated by arrows. The images are filtered to improve clarity; (h) All model hexagons in previous panels (except (g)) are rescaled for comparison and they are well aligned with each other. Hexagon model in (g) is obviously anticlockwise tilted for 30° from other one. It reveals that the graphene domain is twisted with its neighbor at almost 30º. Above images are measured under the same scan angle. The corresponding unfiltered raw images of **e**–**g** are shown in Fig. S8; (i) The boundaries of domains in the top layer of bilayer graphene were unveiled after powerful laser treatment under ambient condition, as the boundaries are more chemically reactive than the pristine graphene lattice.

Under the growth conditions for monolayer graphene, double-layer graphene grains are occasionally seen, especially at increased partial pressure of carbon source CH$_4$. Fig. 4a shows a friction image of the epitaxial bilayer graphene on *h*-BN. The top-layer nucleates about the same defects as the first layer. The smaller size of the top-layer stands for slower growth than the bottom one. The nearly perfect alignment between top- and bottom-layer hexagons implies one particular stacking order. The incoming carbon radicals tend to arrange themselves into thermodynamically stable AB stacking structure during epitaxy.[58] As shown in Figure 4b, clear moiré pattern observed on both terraces exhibits uninterrupted periodicity across the step edge. Atomically resolved AFM images on bottom-, top-layer graphene and *h*-BN substrate (Fig. 4d, 4e, and 4f) confirm that their crystal structures are well aligned with each other. The bilayer graphene is mainly in AB stacking as evidenced by Raman spectrum (See Fig. S11). We noticed that the up-right region between two white dashed lines in Fig. 4a is a graphene domain with lattice 30° -tilted (Fig. 4g), however, a closer look into the moiré pattern showed a continuous pattern across the boundary. The seemingly contradiction can be explained as follows: suppose that the bottom layer graphene is aligned precisely with the h-BN substrate, they will generate a moiré pattern with a periodicity around 14 nm. Then we put an additional graphene layer on top of this moiré pattern with a rotation angle of 30° , we will get another moiré pattern whose periodicity in reciprocal space is the difference in the reciprocal lattice vectors of the top graphene layer and the first moiré pattern. Due to the big difference between the reciprocal lattice vector of the graphene and the first moiré pattern, the wavelength of the second moiré pattern dropped to less than 2 nm regardless its orientation, which is difficult to be imaged by AFM. As a result, only the first moiré pattern is visualized, which overlaps continuously with moiré patterns on other regions across the boundary.

In conclusion, we studied the micro-structural features of single crystal graphene grains grown on insulating *h*-BN with friction AFM. Moiré pattern was observed indicating the van der Waals epitaxial nature of such graphene. From the geometric analysis of moiré pattern and atomic images simultaneously acquired on graphene and substrate, it is determined that the lattice of graphene single crystals aligns precisely with that of *h*-BN within an error less than 0.05 °. The edges of the graphene grains are found to be along armchair direction, consistent with the theoretical prediction for graphene grown on insulating substrates. Besides realizing graphene single crystal grain in micrometer size on *h*-BN, the present work uncovers a crucial issue about the graphene lattice registration on h-BN and might trigger more interests in fundamental as well as application studies with the atomically engineered graphene/h-BN hetero-structure.

While the current manuscript is being reviewed, we notice a related work by metal catalyzed CVD process done by Kim *et al.*, published online[18]. They claim turbo-static stacking of graphene/h-BN/Cu and aligned h-BN/graphene/Cu with angular separation between graphene and h-BN smaller than 1 degree. We are updated that the very recent paper by Yang *et al.*,[59] grow epitaxial graphene on *h*-BN by plasma enhanced CVD after the first round review of our manuscript.

**Methods:** Graphene was deposited on the h-BN via chemical vapor deposition. Graphene samples were grown at 1200 °C by flowing CH$_4$:H$_2$ at 5 : 5 s.c.c.m. for 60–300 minutes with pressure below 12 mbar. After growth, samples are cooled down to room temperature by keeping the argon flowing. The morphology, grain size, shape and crystallographic orientation of the CVD graphene can be recorded in contact mode (Veeco NanoScope V AFM). To obtain a high accuracy, calibration in atomic resolution was performed with newly cleaved highly ordered pyrolytic graphite (HOPG) before measurement. Parameters like integral gain, set-point and scan rate were adjusted to get high quality images. Several hours pre-scanning were carried out to warm up the scanner in order to obtain a high stability in imaging. See supplementary information for the details about fabrication, characterization and transport measurement of the graphene/h-BN hetero-structure.

**Acknowledgements:** We thank Z.X. Shen and Y. Cui from Stanford University for fruitful discussions. We acknowledge B. Gao for his support in transport measurement trials. The work in Shanghai Institute of Microsystem and Information Technology, Chinese Academy of Science is partially supported by the National Science and Technology Major Projects of china (Grant No. 2011ZX02707), Chinese Academy of Sciences (Grant No. KGZD-EW-303), CAS International Collaboration and Innovation Program on High Mobility Materials Engineering, the National Natural Science Foundation of China (Grant No. 11104303, 11274333, 11204339 and 61136005), and the projects from Science and Technology Commission of Shanghai Municipality (Grant No. 12JC1410100 and 12JC1403900). Single crystal flakes from Momentive Performance Materials, Inc., USA is highly appreciated.

**Author contributions:** M. J. and X. X. directed the research work. H. W., X. X. and S. T. conceived and designed the experiments. S. T. fabricated the graphene samples. S. T. and Y. Z. performed the AFM experiments with support from T. L., L. L. and H. W.. H. X. and X. L. fabricated the electronic devices. H. W. carried the transport measurements. H. W., A. L. and S. T. analyzed the data and designed the figures. H. W. performed the theoretical calculations. H. W., A. L, F. H., X. X. and M. J. co-wrote the manuscript and all authors contributed to the critical discussions of the manuscript.

**Additional information:** Supplementary information accompanies this paper including: Description of graphene growth on h-BN, details about AFM measurement, calibration on graphene and h-BN surface, Raman spectroscopy and transport characterization. This material is available free of charge via the Internet at http://.

**Competing financial interests:** The authors declare no competing financial interests.

# Supplementary Information:

## Precisely Aligned Graphene Grown on Hexagonal Boron Nitride by Catalyst Free Chemical Vapor Deposition

Shujie Tang,[1,] Haomin Wang,[1, a)] Yu Zhang,[2] Ang Li,[1] Hong Xie,[1] Xiaoyu Liu,[1] Lianqing Liu,[2] Tianxin Li,[3] Fuqiang Huang,[4] Xiaoming Xie,[1,a)] Mianheng Jiang[1]

[1] State Key Laboratory of Functional Materials for Informatics, Shanghai Institute of Microsystem and Information Technology, Chinese Academy of Sciences, 865 Changning Road, Shanghai 200050, P.R. China
[2] State Key Laboratory of Robotics, Shenyang Institute of Automation, Chinese Academy of Sciences,114 Nanta Street, Shenhe District, Shenyang 110016, P.R. China
[3] National Laboratory for Infrared Physics, Shanghai Institute of Technical Physics, Chinese Academy of Sciences, 500 Yu Tian Road, Shanghai 200083, P.R.China
[4] CAS Key Laboratory of Materials for Energy Conversion, Shanghai Institute of Ceramics, Chinese Academy of Sciences, Shanghai, 200050, P.R. China
a) Correspondence and requests for materials should be addressed to: hmwang@mail.sim.ac.cn, xmxie@mail.sim.ac.cn

**List of the content:**
1) Details for graphene growth
2) Details for scanning probe studies
3) Calibration on graphene, graphite and h-BN surface for atomic-resolution friction measurement
4) The error for lattice orientation measurement
5) Determining the wavelength and orientation of moiré patterns
6) The corresponding unfiltered raw images of Fig. 1 e–g
7) The wavelength and rotation of moiré pattern
8) The corresponding unfiltered raw images of Fig. 3
9) The corresponding unfiltered raw images of Fig. 4.
10) Thickness determination and Raman spectroscopy
11) Unit cell in two layer graphene
12) The possibility of perfectly sewing for the epitaxial graphene grains
13) Transport characterization on epitaxial graphene/h-BN devices

## 1) Details of graphene growth

Commercially available h-BN was exfoliated on $SiO_2$ substrates; the h-BN flakes have a large variation in their lateral size and thickness. The size of graphene varies from several to tens of micrometers. The thickness ranges from 10 to 50 nm. Graphene was deposited over the h-BN via CVD method. Firstly, we annealed the h-BN flakes on $SiO_2$/Si at 1200 °C at low pressure with a continuous argon flow of 50 standard cubic centimeters per minute (s.c.c.m.) for 30 minutes. We then grew the graphene at 1200 °C by flowing $CH_4$:$H_2$ at 5 : 5 s.c.c.m. for 60–300 minutes with pressure below 12 mbar. Samples are cooled for ~100 min while the argon flow is maintained. The morphology, grain size, shape and crystallographic orientation of the CVD graphene can be controlled by varying growth condition. The polycrystalline graphene is often seen in the sample grown at a lower temperature during the formation of a monolayer graphene while an increase in partial pressure of $CH_4$ could enable the growth of the second layer on the first layer graphene. Fig. S1 shows one of the largest single crystal graphene we obtained, its' cater-cornered length is about 2.5 μm.

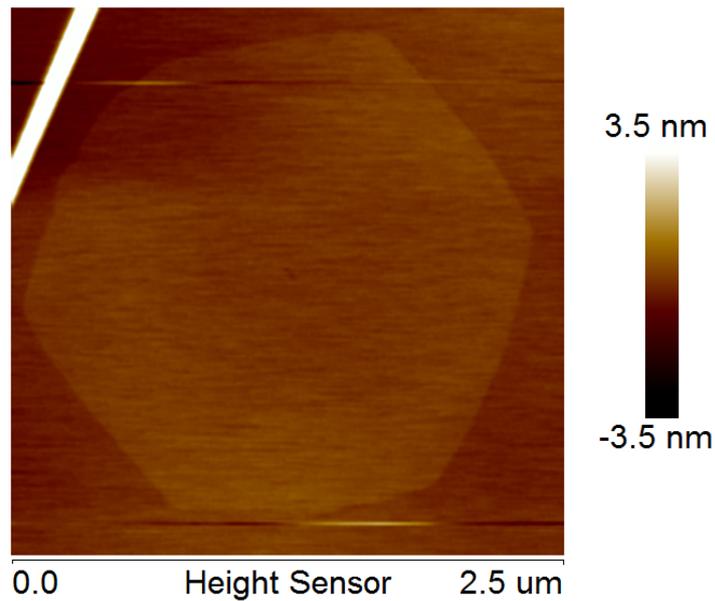

**Figure S1.** A single crystal graphene grown on hBN via CVD

**2) Details of scanning probe studies**

AFM images on different surfaces were recorded in contact mode (Veeco NanoScope V AFM) using SNL-10 AFM tips having nominal tip radius of less than 10 nm. These cantilevers have force constant k = 0.05 -- 0.5 N/m. To obtain a high accuracy, scanners with a travel range less than 10μm along X and Y direction was used. Calibration in atomic resolution was performed with newly cleaved highly ordered pyrolytic graphite (HOPG) before measurement. After calibration the mean distance between vicinal carbon atoms is 0.142nm. Integral gain and set-point were adjusted as low as possible to get good image. The scan rate is set to a value in the range of 10-60 Hz to reduce the noise from thermal drift. Several hours pre-scanning were carried out to warm up the scanner in order to obtain a high stability in imaging.

**3) Calibration on graphene, graphite and h-BN surface for atomic-resolution friction measurement**

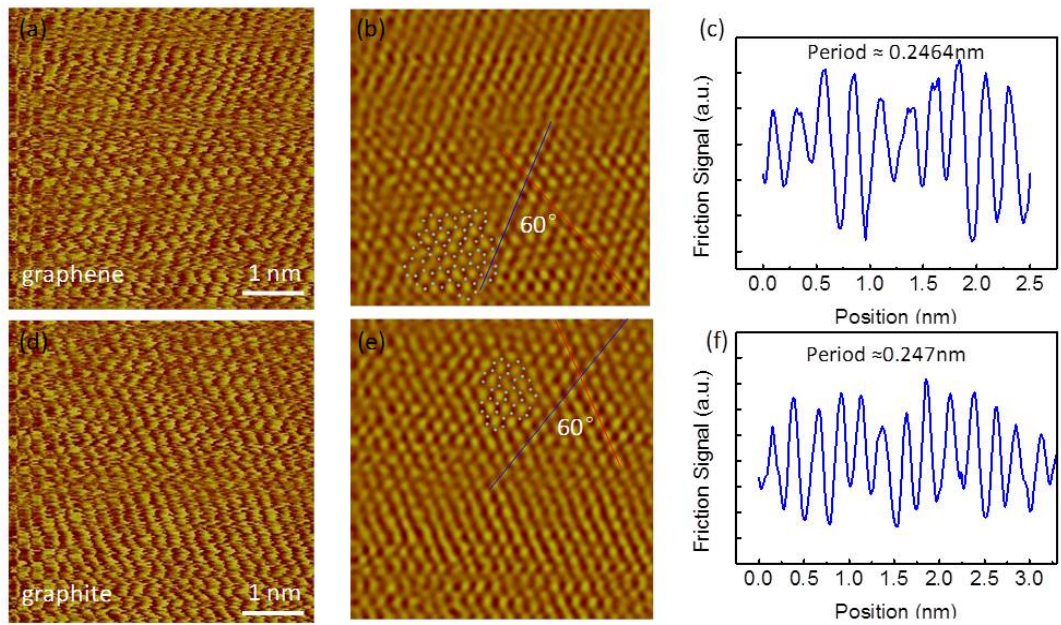

**Figure S2.** Atomic-resolution friction measurement on graphene and graphite for calibration. (a) Unfiltered and (b) low-pass filtered friction AFM image of the crystal lattice in graphene monolayer deposited on top of an oxidized Si wafer. For graphene, the raw images of friction show that the tip moves slightly with stick-slip motion. The filtered images give clearer periodic sites of the friction force signal. Some azure dots are superimposed on the images to exhibit the graphene honeycomb lattice; (c) Friction force signal as a function of distance on sample areas along the blue line shown in (b), the friction force signal demonstrates a period of about 0.2464nm; (d) Raw and (e) low-pass filtered friction AFM image on a fresh graphite surface. The azure dots are superimposed on the filtered images to exhibit the lattice; (f) Friction force signal as a function of distance on sample areas along the blue line shown in (b), the friction force signal demonstrates a period of about 0.247nm.

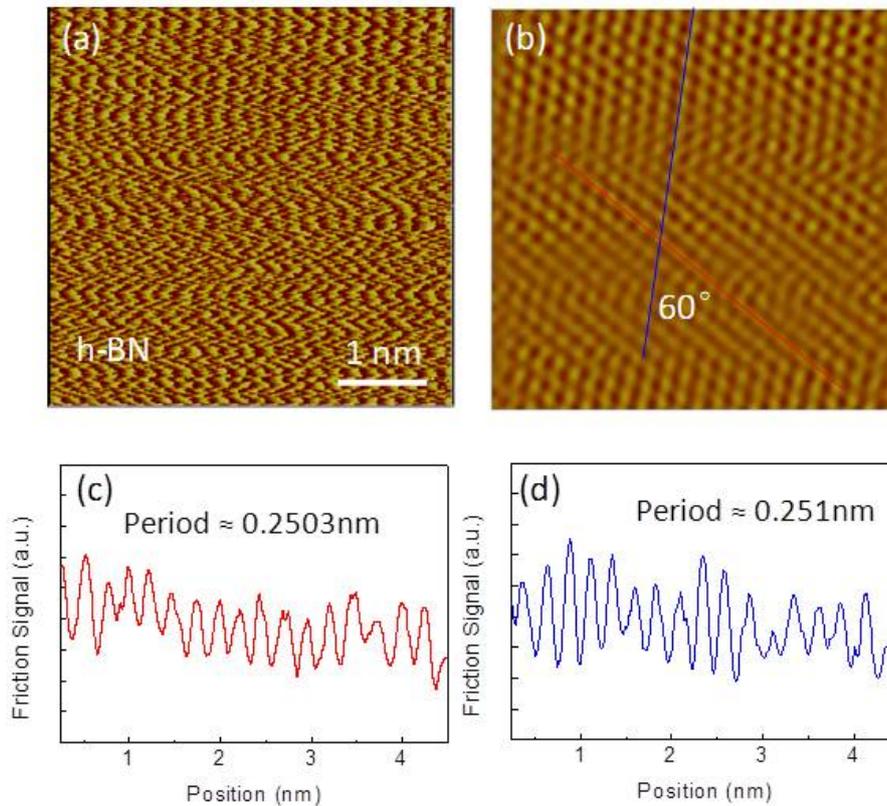

**Figure S3.** Atomic-resolution friction measurement on h-BN for calibration. (a) Unfiltered and (b) low-pass filtered friction AFM image of the crystal lattice in h-BN. For h-BN, The raw images of friction show that the tip moves slightly with stick-slip motion. The filtered images give clearer periodic sites of the friction force signal. Some azure dots are superimposed on the images to exhibit the h-BN honeycomb lattice; (c) and (d) friction force signal as a function of distance on sample areas along the blue and red lines shown in (b), respectively. The friction force signal demonstrates a period of about 0.251nm.

4) **The error for lattice orientation measurement**

To get relative lattice orientations, we referred to atomic resolution images and their corresponding FFT patterns. The measured angles in between lattice vectors are record over the graphene grains. We recorded near 300 data points on a total of 18 different graphene grains. Error in this orientation measurement varies by data points and is typically ±2º, sometimes to ±3º.

5) **Determining the wavelength and orientation of Moiré patterns**

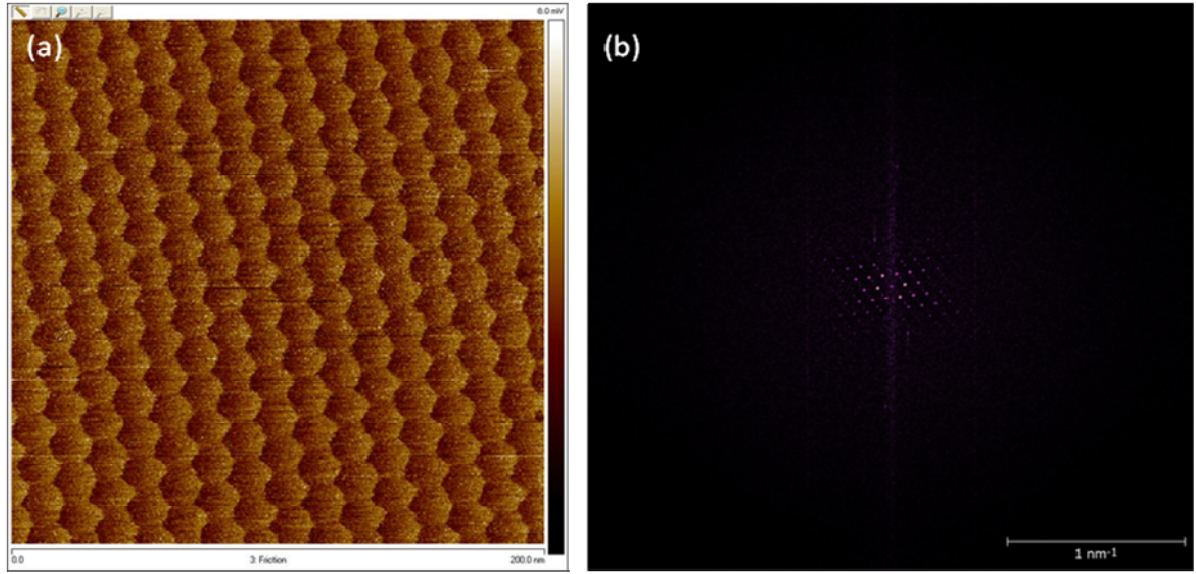

**Figure S4** Real space and Fourier transforms of moiré patterns (a) Lattice showing a moiré pattern produced by graphene on h-BN; (b) Fourier transform of (a) showing the six graphene lattice points near the center of the image. The visualization and analysis software we use is Gwyddion from http://gwyddion.net/. The average wavelength of the moiré pattern is about 14nm. The orientation of the moiré patterns was determined from the spectrum after Fourier transform. The orientation of real lattice can be obtained by rotate that of reciprocal lattice with 30° anticlockwise.

**6) The corresponding unfiltered raw images of Fig. 1 e–g**

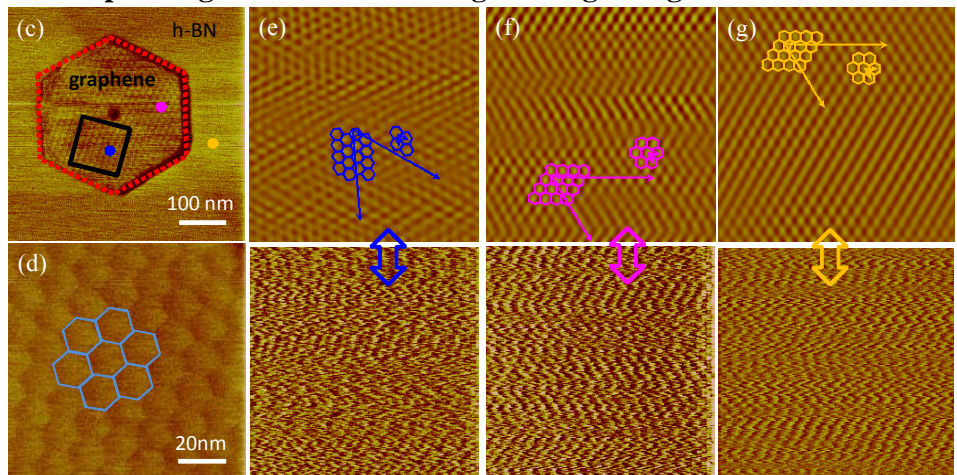

**Figure S5** Determination of the rotational orientation of graphene sheets with respect to h-BN. (c) AFM friction image of a hexagonal graphene single crystal on h-BN. (d) a closer view (100 × 100 nm$^2$) of the giant lattice in the black box of (c), the giant lattice exhibits a hexagonal symmetry with a lattice constant of about 14 nm; The friction images (e), (f) and (g) showing the atomic lattice taken from the blue, pink and orange dots in panel(c) in an area of (5 × 5 nm$^2$), the corresponding unfiltered raw images of **e–g** are shown.

## 7) The wavelength and rotation of the moiré pattern

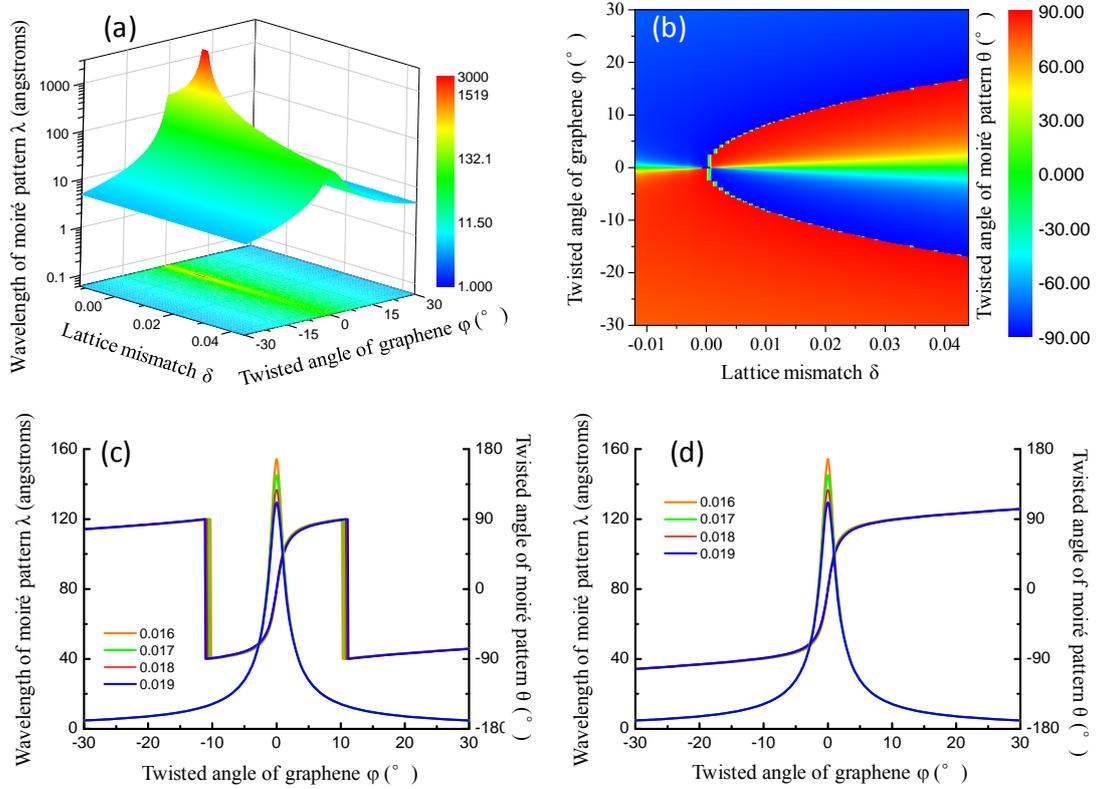

**Figure S6** (a) The wavelength of moiré pattern as a function on lattice mismatch of graphene δ and twisted angle of graphene with respect to h-BN $\phi$; (b) The twisted angle of moiré pattern as the function of lattice mismatch and twisted angle of graphene; Moiré pattern wavelength and rotation as a function of the angle between the graphene and h-BN in different lattice mismatch (c) after and (d) before 90° subtraction.

The wavelength of moiré pattern as a function on lattice mismatch of graphene and twisted angle of graphene with respect to h-BN is plotted in Fig. S6a. Owing to six-fold symmetry in the lattice of both graphene and h-BN, moiré pattern from graphene/h-BN stacking produces a periodicity of 60º in their twisted angle. In this plot, the twisted angle varies from -30º to 30º. The lattice mismatch δ is assumed varying from -1.2% to 4.4%. It is found that the wavelength of moiré pattern decreases with the increase of the absolute value of δ. Moiré patterns for all orientations of graphene on h-BN have a maximum possible length when the twisted angle of graphene $\varphi$ equals zero. In other words, the maximum moiré pattern can be observed when the lattice orientation of graphene grains is precisely aligned with the underlying h-BN. In real case, δ is only in the range of 1.6% - 1.9%, the corresponding wavelength of moiré $\lambda$ equals to about 13 ~15nm. In addition, the twisted angle of moiré pattern as the function of lattice mismatch and twisted angle of graphene is plotted in Fig. S6b. When the twisted angle of the moiré pattern goes beyond the value of 90 °, it is subtracted with 180 ° for clarity in the plot of Fig. S6b. And similar treatment was done for the case of rotation θ lower than -90 °. Figure S6c is the final plot for the process while the Figure S6d is the original plot.

**8) The corresponding unfiltered raw images of Fig. 3.**

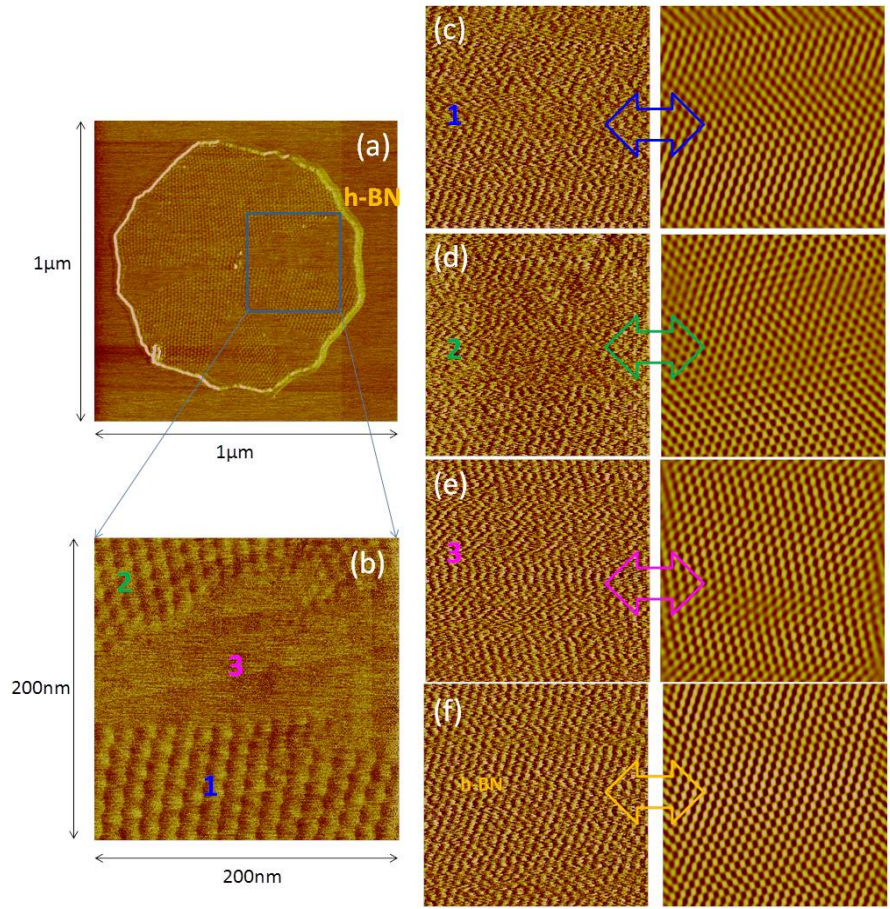

**Figure S7** Investigation of the rotational orientation of polycrystalline monolayer graphene on h-BN. (a) A typical polycrystalline monolayer graphene grown on h-BN surface. (b) Magnified friction images from the black box in panel (a) is shown. The friction images (c), (d) and (e) showing the (5 nm ×5 nm) atomic lattice taken from the blue, pink, green and orange positions in panel(a) and (b). The filtered images and the corresponding unfiltered raw images are shown respectively.

**9) The corresponding unfiltered raw images of Fig. 4**

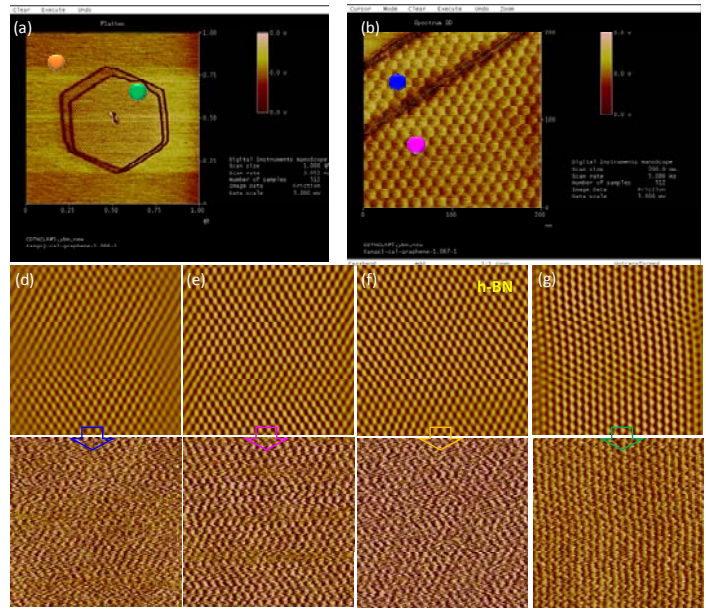

**Figure S8** Investigation of the rotational orientation of bilayer graphene grown on h-BN. (a) A typical topography of a bilayer graphene grown on h-BN surface. (b) A zoom in view of moiré pattern on the two-layer graphene; The friction images (d), (e), (f) and (g) showing the (5 nm ×5 nm) atomic lattices taken from the blue, pink, orange and green dots in panel(a), (b), The filtered images and their corresponding unfiltered raw images are shown, respectively.

**10) Thickness determination and Raman spectroscopy**

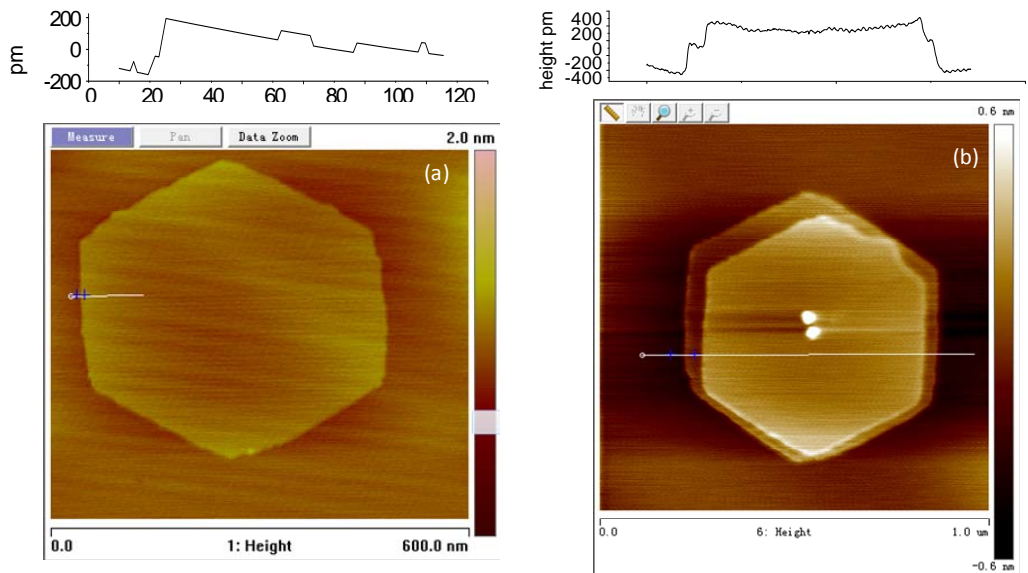

**Figure S9.** Height characteristics of graphene (a) monolayer and (b) its bilayer on hBN

As shown in Fig. S9a, the height of 0.345nm confirm that the graphene is monolayer. The topography of the graphene flakes shown in Fig. S9b tells that its thickness is about 0.7nm. It is obvious that the graphene flake is mainly of bilayer, although the top layer is slightly smaller than the bottom one.

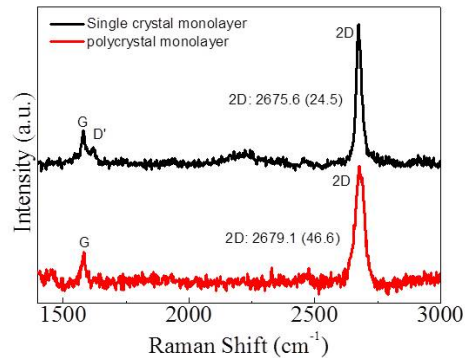

**Figure S10.** Raman spectrum of graphene monolayer in single crystal and in polycrystal, showing the main Raman features the G, and 2D bands taken with a laser line of 532nm. The first order Raman allows G band and the very strong 2D band of monolayer graphene which is described by one Lorentzian and a peak frequency of 2675cm$^{-1}$ with FWMH of 24.5cm$^{-1}$ for single crystal and 2679cm$^{-1}$ with FWMH of 46.6cm$^{-1}$.

Raman spectra are obtained with a thermo-fisher micro-Raman instrument with a exciting laser of 532nm. The spot of laser is focus to about 1μm in diameter. The power of laser keeps less than 3mW. Raman spectroscopy was used to support the conclusion that the polycrystal exist in the graphene grain. The Raman spectra of two graphene grains are shown in Fig. S10, where the background is subtracted. The doublet at about 1590cm$^{-1}$ is from the superposition of the G and D'. The former is a long wavelength optical photon while the later is the critical point in the optical phonon dispersion arising from relaxation of wave vector conservation rules induced by defects.[1,2] It is found that the D peak (1360cm$^{-1}$ wave-number) is associated with the defects in the graphene lattice. The high ratio of 2D peak to G peak confirms the graphene grain in monolayer thickness. Fig. S10 shows that Raman spectrum of the graphene flakes used in the study, black curve shows the Raman spectrum of single crystal graphene in hexagonal shape presented in the Fig. 1C. The narrow width of 2D peak (24.5cm$^{-1}$) spectrum was identified to determine the single crystalline in graphene grain. Red curve shows the boarder width of 2D peak (46.6cm$^{-1}$) for the poly-crystal flake presented in the Figure 3, and nearby flake contained many single crystalline patches.

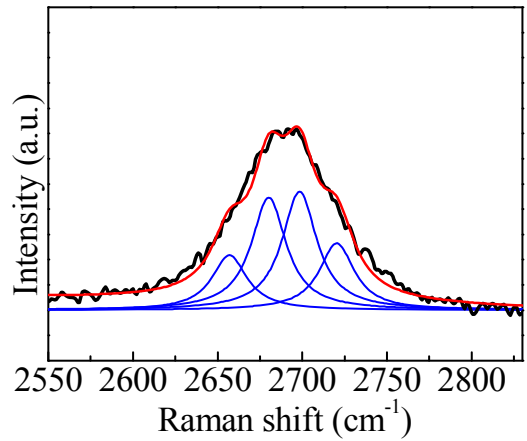

**Figure S11.** The Raman spectrum of bilayer graphene excited with 532nm laser line. The four fitted Lorentzians are also shown.

The number of graphene layer is determined from the line-shape analysis of the second-order 2D band. The 2D peak shown in Fig. S11 can be well fitted by four Lorentzian peaks. The four fitted Lorentzian peaks indicate the four double resonance processes which are the fingerprint of bilayer graphene in AB stacking.[3] Although there are some monolayer components and twisted two layer graphene, the AB stacking features are dominating the Raman spectrum of bilayer graphene.

## 11) Unit cell in two layer graphene

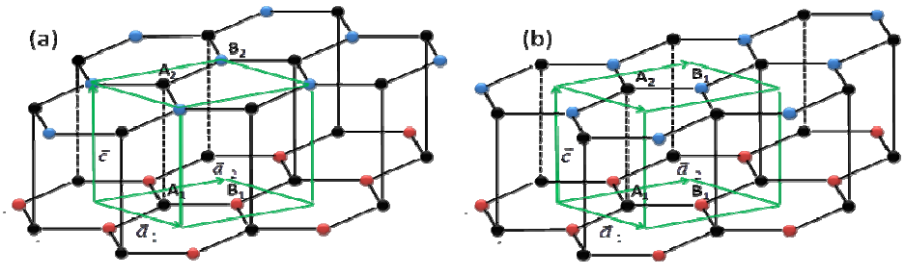

**Figure S12** Lattice structure of a graphene bilayer in a) AA and b) AB stacking. The unit cell is a green parallelepiped

It is clear that the lattice vectors of AB and AA stacking in the graphene are the same as the monolayer graphene in graphene lattice plane. As the AA stacking is very unstable status during the formation, AB stacking is more likely to form in chemical vapor deposition.

## 12) The possibility of perfectly sewing for the epitaxial graphene grains

Here it is necessary to discuss the possibility that the epitaxial graphene grains sew up at the boundary with the perfect structure of honeycomb $sp^2$ carbon. One may expect that the hexagonal carbon lattice in neighboring graphene can meet each other exactly with the perfect sp2 carbon structure. Although the orientations of graphene grains are the same for all of the hexagonal grains, unavoidable sliding of the lattice along the grain boundary could introduce the boundary with lattice distortion or defective sites. Therefore, the epitaxial graphene grains may not sew up at the boundary in perfect honeycomb $sp^2$ structure.

## 13) Transport characterization on epitaxial graphene/h-BN devices

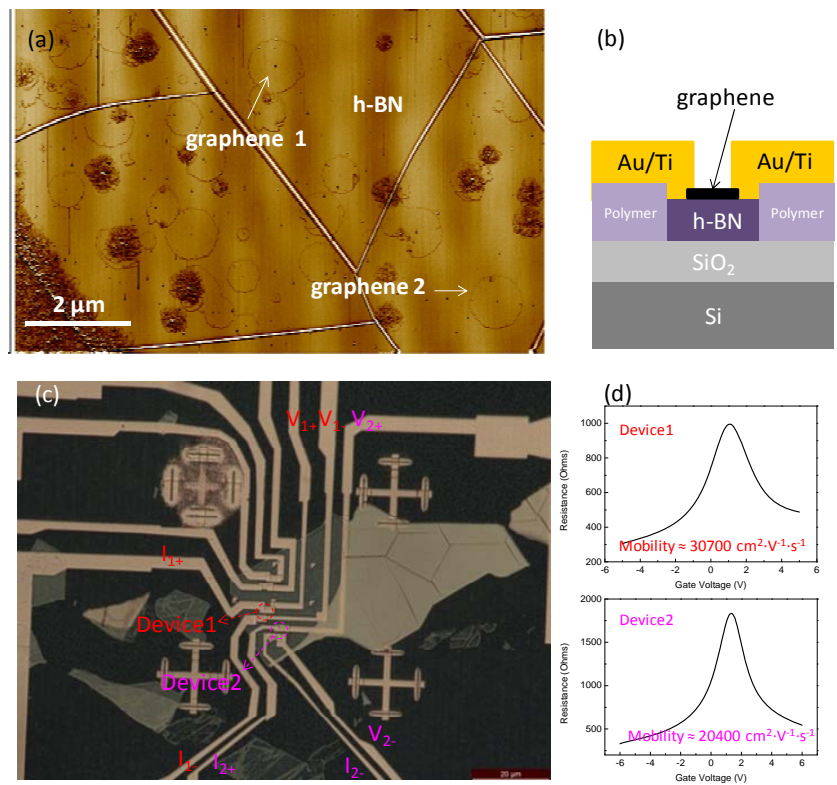

**Figure S13** Transport characterization on epitaxial graphene. (a) AFM topographic image of polycrystalline graphene flakes grown on h-BN. Scale bar 2 μm. (b) Graphene devices structure, designed for transport measurement. (c) Optical micrograph of graphene devices grown on a 10 nm thick hBN crystal. Scale bar 20 μm. (d) Resistance versus back gate voltage for two typical monolayer graphene/h-BN devices. The devices were measured under ambient condition. The resistivity peak is extremely narrow and typically occurs at nearly zero gate voltage.

For the measurement of transport properties, electrical leads are deposited using standard electron beam lithography and lift-off process. The transport properties of the devices are measured under a low driving current 1μA by using low-frequency lock-in technique. Beside the noise suppression, another merit of the a.c. measurement is to reduce the effective impedance at the graphene/metal contact, the capacitance components at the contact may greatly affect the results from d.c. measurement. The gate voltage is applied by using Keithley 2400 sourcemeter. As the thickness of the h-BN flake on 300 nm SiO$_2$/Si substrates is about 10 nm, the effective capacitance C$_g$ can be estimated at about 11.5 nF·cm$^{-2}$. The channel width and length of graphene were 900 nm and 300 nm, respectively. Figure S13d shows the back gate dependence of resistance of the two devices. The field effect mobility (μ) of graphene sheets and the contact resistance can be extracted from formula:

$$R_{tatal} = \frac{L/W}{\mu\sqrt{(n_0 e)^2 + (V_g - V_{dirac})^2 C_g^2}} + R_c$$

Where resistance $R_{total}$ includes the channel resistance of graphene and contact resistance $R_c$ between graphene and the Au/Ti metal, "e" represents elementary charge, and $n_o$ is the residual carrier density induced by charge impurities.[4] L and W indicate channel length and width, respectively. The gate dependence of $R_{total}$ was measured at room temperature. It is found that the sample is slightly p-doped. It may be due to the proximity effect from the metal contact and water absorption.[5] The field effect mobilities extracted from the two graphene samples are about 20400 and 30700 cm$^2$·V$^{-1}$·s$^{-1}$. Contact resistances are about 177 and 170 Ohms. The high mobility indicates that graphene grown on h-BN dielectrics has a strong potential in the nano-electronics.